\documentclass[aps,groupedaddress,showpacs,letterpaper,twocolumn]{revtex4}
\usepackage{graphicx} 
\usepackage{amsmath}

\begin{document}

\title{Efficient multiparticle entanglement via asymmetric Rydberg blockade }
\author{ M. Saffman}
\affiliation{
Department of Physics,  1150 University Avenue,
University of Wisconsin,  Madison, Wisconsin 53706, USA
}
\author{K. M\o{}lmer }
\affiliation{Lundbeck Foundation Theoretical Center for Quantum System Research, Department of Physics and Astronomy,
University of Aarhus, DK-8000 \AA{}rhus C, Denmark}
 \date{\today}

\begin{abstract}
We present an efficient  method for producing $N$ particle entangled states using Rydberg blockade interactions. Optical excitation of Rydberg states that interact weakly, yet have a strong coupling to a second control state is used to achieve state dependent qubit rotations in small ensembles. On the basis of quantitative  calculations we predict that an $N=8$ Schr\"odinger cat state can be produced with a fidelity of 84\%  in  cold Rb  atoms.
\end{abstract}

\pacs{03.67.Bg, 32.80.Qk, 32.80.Ee}
\maketitle

Entanglement lies at the heart of quantum information processing and is also a valuable resource for extending precision measurements beyond bounds set by classical statistics. Recent years have seen a steady progression towards entanglement of larger and larger objects.
Although macroscopic ensembles have been successfully  entangled\cite{Julsgaard2001}, the entanglement achieved per atom was very
low. Maximally entangled cat states of  six atoms, as well as ``W" states of eight atoms  have been achieved in groundbreaking experiments with cold ions\cite{Leibfried2005,Haffner2005}.  In this letter we introduce an efficient technique for generating maximally entangled states which is applicable to any system which supports asymmetric state dependent blockade interactions. We give quantitative estimates for the preparation fidelity for entanglement of the clock states of Rb atoms using Rydberg blockade, which may enable  improvement in the accuracy of an atomic clock.

Consider the situation shown in Fig. 1 where $N$ atomic quits, each with basis states $|0\rangle, |1\rangle$, are confined in a volume $V.$ We assume  states $|0\rangle, |1\rangle$ are weakly
interacting over time scales of interest but can be transferred to additional interacting states $|s\rangle, |p\rangle$. Single particle excitations of $|s\rangle $ are allowed but there is a large energy gap $U_{\rm ss}=\hbar\Delta_{\rm ss}$  which blocks two-particle excitations.  States $|s\rangle, |p\rangle$ are also strongly interacting with a large gap $U_{\rm sp}=\hbar\Delta_{\rm sp}$, however states $|p\rangle$ interact weakly with each other so that the two-particle interaction energy $U_{\rm pp}=\hbar\Delta_{\rm pp}$ satisfies $\Delta_{\rm pp}\ll \Delta_{\rm sp},\Delta_{\rm ss}.$

\begin{figure}[!t]
\centering
\includegraphics[width=8.5cm]{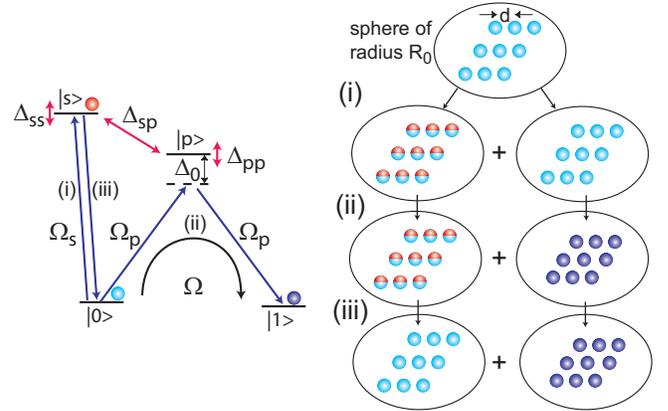}
\vspace{-.3cm}
\caption{(color online) Level scheme (left) and sequence of operations for cat state generation (right).
$\Omega$ is the effective Rabi frequency coupling states $|0\rangle, |1\rangle$.} 
\label{fig.1}
\end{figure}

With the above resources $N$ atom entangled states can be synthesized in a few interaction steps by the following protocol. We first prepare the $N$ atom product state $|\psi\rangle = |0,0,...0\rangle$. The ground state $|0\rangle$ is coupled to $|s\rangle$ with an interaction Hamiltonian ${\mathcal H}_1$ such that the Rabi frequency (from now on we put $\hbar=1$), given by $\Omega_{\rm s}/2=\langle s |{\mathcal H}_1| 0 \rangle $, satisfies $|\Omega_{\rm s}|\ll \Delta_{\rm ss}$ . In step (i) we apply
$\mathcal H_1$ to all atoms for a time $ t_1 =\pi/(2\sqrt N |\Omega_{\rm s}|)$ to create the entangled state
\begin{equation}
|\psi\rangle =\frac{1}{\sqrt2}\left(\frac{1}{\sqrt N}\sum_{j=1}^N |0,0,s^{(j)}...0\rangle + |0,0,...0\rangle\right).
\label{eq.psi1}
\end{equation}
We then  invoke  a second interaction Hamiltonian
${\mathcal H}_2={\mathcal H}_{20}+ {\mathcal H}_{21}$ with  corresponding
Rabi frequencies
$\Omega_{\rm p0}/2=\langle p |{\mathcal H}_{20}| 0 \rangle $,
$\Omega_{\rm p1}/2=\langle p |{\mathcal H}_{21}| 1 \rangle $, and the same detuning $\Delta_0$ on both transitions, see Fig. 1.
For simplicity we will assume $\Omega_{\rm p0}=\Omega_{\rm p1}=\Omega_{\rm p}=|\Omega_{\rm p}|.$
After a definite interaction time of $t_2=\sqrt{2}\pi/\Omega_{\rm p}$  in the resonant $(\Delta_0=0)$ case, and $t_2=2\pi \Delta_0/\Omega_{\rm p}^2$ in the non-resonant $( \Delta_0\gg \Omega_p)$ case, ${\mathcal H}_2$ induces a transfer from $|0\rangle$ to $|1\rangle$ in all the atoms via the Rydberg state $|p\rangle$, unless this process  is blocked by population in the Rydberg $|s\rangle$ state.
In the limit where $\Delta_{\rm pp}\ll \Omega\ll \Delta_{\rm sp}$
step (ii)  transforms (\ref{eq.psi1})  into
\begin{equation}
|\psi\rangle =\frac{1}{\sqrt2}\left(\frac{1}{\sqrt N}\sum_{j=1}^N |0,0,s^{(j)}...0\rangle+|1,1,...1\rangle\right).
\label{eq.psi2}
\end{equation}
We finish in step (iii) by applying $-{\mathcal H}_1$ for a time $2 t_1$ to reverse the first excitation step,  giving 
\begin{equation}
|\psi\rangle =\frac{1}{\sqrt2}\left(|0,0,...0\rangle+|1,1,...1\rangle\right)
\label{eq.psi3}
\end{equation}
which is a $N$ atom maximally entangled state. We see that, independent of $N$, only three preparation steps are needed.
We show below that the requirement
of strong and state dependent asymmetric couplings may be satisfied
by dipole-dipole interactions of Rydberg atoms\cite{Jaksch2000,Lukin2001}.  Rydberg blockade effects have now been observed in a number of experiments in both many-body\cite{Rydbergmeso} and single atom\cite{Rydbergsingle} settings. The Rydberg blockade has been suggested previously as a route to multi-particle entanglement\cite{Unanyan2002,Moller2008} and recent work from M\"uller et al.\cite{Muller2008} is based on ideas closely related to those presented here.  There are, however, significant differences including  our use of blockade in step (i) of the above protocol which removes the need for separately addressing a control atom. This allows all atoms to reside in one ensemble which provides a better geometrical scaling of the interactions.

The  fidelity with which state (\ref{eq.psi3}) can be prepared in an experiment depends on the strength of the blockade interactions and the degree to which the couplings are asymmetric.
We proceed by estimating the effective Rydberg interaction strengths $\Delta_{\rm sp}, \Delta_{\rm pp}$.
The interaction between  atoms in the ``control" state $|s\rangle$ and the ``target" state $|p\rangle$ is of resonant dipole nature between two-atom
states $|sp\rangle,|ps\rangle$. The interaction between classical dipoles is anisotropic and has a zero when the angle between the dipoles is $\theta=\cos^{-1}(1/\sqrt3).$ This anisotropy would lead to unacceptable errors in the present setting. However, for small external fields  the atomic Zeeman states are degenerate  and the interaction couples states with different $m_s, m_p$ quantum numbers. This leads to a finite interaction strength $\Delta_{\rm sp}$ at all angles. It can be shown that the behavior corresponding to interaction of classical dipoles with angular zeroes is recovered by applying a large magnetic field that selects a single pair of Zeeman states. The resonant interaction has a $1/R^3$ scaling so
we can write $\Delta_{\rm sp}(R)=\Delta_{\rm sp}(d) (d/R)^3$ where $d$ is a characteristic length scale that we will set equal to the smallest interatomic separation $d.$
The target-target interaction $\Delta_{\rm pp}$ is due to a F\"orster process\cite{Forster1949}. We will choose states and values of $d$  such that we are working in the van der Waals limit of this interaction which gives the distance scaling $\Delta_{\rm pp}(R)=\Delta_{\rm pp}(d) (d/R)^6.$ The energy shift of each atom is thus dominated by its nearest neighbors in the ensemble.

Since $\Delta_{\rm sp}\sim 1/R^3$ and $\Delta_{\rm pp}\sim 1/R^6 $ the condition of strongly asymmetric Rydberg interactions can be readily met by
choosing $R$ sufficiently large. The asymmetry is maximized for small $n$ since the resonant dipole allowed interaction between $|s\rangle, |p\rangle$ scales as $\Delta_{\rm sp}\sim n^4$ while the second order F\"orster process leading to $\Delta_{\rm pp}$ scales as
$\Delta_{\rm pp}\sim n^{11}.$
The lower limit on $n$ is  set by the blackbody limited spontaneous emission lifetime $\tau_{\rm p}\sim n^2 .$

\begin{figure}[!t]
\centering
\includegraphics[width=8.cm]{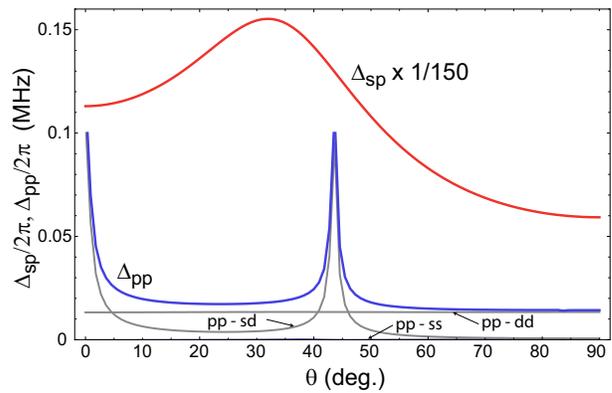}
\vspace{-.4cm}
\caption{(color online) Calculated blockade shifts as a function of the angle of the molecular axis for $^{87}$Rb,  $|s\rangle=|41s_{1/2},m=1/2\rangle,$
$|p\rangle=|40p_{3/2},m=1/2\rangle$ with magnetic field $B=10^{-7}~\rm T$ at $d=3~\mu\rm m$.  The interaction $\Delta_{\rm pp}$ includes contributions from the indicated
F\"orster channels. }
\label{fig.2}
\end{figure}

We have searched for parameters in the small $n$ regime that minimize the error in creation of state (\ref{eq.psi3}) by performing extensive numerical studies of the interaction between different Rydberg states in Rb drawing on the exposition of the
dominant F\"orster channels given in \cite{Walker2008}.
As shown in Fig. \ref{fig.2} we find  large interaction asymmetries using $n_s=41$ $s_{1/2}$ states for 
$|s\rangle$ and
$n_p=40$ $p_{3/2}$ states  for 
$|p\rangle$ at a lattice spacing of $d=3.~\mu\rm m$.
  In calculating the $|p\rangle$ state interactions $\Delta_{\rm pp}$ we have included all channels of the form
$npnp\rightarrow n'sn''s,  n'sn''d,  n'dn''d$.   We see from the figure that $\Delta_{\rm sp}/\Delta_{\rm pp} > 150$  for all angles . In addition to multiparticle entanglement
this large asymmetry will also facilitate implementation of a three-bit Toffoli gate\cite{Brion2007}.

We consider an implementation with a cubic lattice of spacing $d$ occupied by one atom per site inside a sphere of radius $R_0.$ A protocol for preparing a lattice with this type of  spatially localized occupation was described by us recently in Ref. \cite{Saffman2008}.
The angle dependent peaks in $\Delta_{\rm pp}$ are not of particular concern since the cubic lattice can be oriented to avoid the corresponding angles. For simplicity we
have characterized the angle averaged interactions by $\bar \Delta_{\rm sp}(d)=\int_0^{\pi/2} d\theta \, \Delta_{\rm sp}(d,\theta)\sin(\theta)$ and similarly for $\bar\Delta_{\rm pp}. $  Performing the integrations we find
$(\bar\Delta_{\rm sp}, \bar\Delta_{\rm pp})/2\pi = (14.4, 0.019)~\rm MHz.$

In order for step (i) of the entanglement protocol to be effective it is also necessary that the  $|n_s=41,s_{1/2}\rangle$ states exert a strong blockade over the entire sphere of radius $R_0.$ At $d=3~\mu\rm m $ we find $\Delta_{\rm ss}(d)=3.7~\rm MHz$ and the interaction is essentially isotropic\cite{Walker2008}. This is larger than $\Delta_{\rm pp}$ but the interaction strength decreases as $1/R^6$ and is insufficient for strong blockade over a sphere with $R_0>d.$ We note that this difficulty can be readily solved as follows. The first Rydberg excitation step (i) is made to a level $|s'\rangle$ that has a large value of $n$ and provides strong blockade over the entire ensemble. The level $|s'\rangle$ is then transferred
to $|s\rangle$  using a two-photon transition which prepares the state of Eq. (\ref{eq.psi1}) even though $\Delta_{\rm ss}$ may be small.
The additional transfer steps are then run backwards in step (iii).
The possibility of performing these additional steps allows us to ignore the small errors associated with blockade of the control state.

To examine the fidelity of our scheme it is instructive to recall how the errors scale  for a two-atom Rydberg blockade phase gate. A phase gate between a control atom $(c)$ and target atom $(t)$ involves the steps\cite{Jaksch2000}: i) $\pi$ pulse  $|1\rangle_{\rm c}\rightarrow i|r\rangle_{\rm c}$, ii) $2\pi$ pulse  $|1\rangle_{\rm t}\rightarrow i|r\rangle_{\rm t}\rightarrow -|1\rangle_{\rm t}$, iii) $\pi$ pulse $i|r\rangle_{\rm c}\rightarrow -|1\rangle_{\rm c}$.
Assuming  ground  to Rydberg state oscillations can be driven with high accuracy, as has been demonstrated in recent experiments\cite{Johnson2008}, the dominant errors come from imperfect blockade in step ii) with error $E_{\rm bl}\sim \Omega^2/\Delta^2$ ($\Delta$ is the dipole-dipole interaction shift) and spontaneous emission of the control atom with error $E_{\rm se}\sim 1/\Omega\tau $ where $\tau$ is the Rydberg state spontaneous lifetime. The sum of the two errors is minimized for $\Omega\sim \Delta^{2/3}/\tau^{1/3}$ which leads to a gate error that scales as\cite{Saffman2005} $E=E_{\rm bl}+E_{\rm se}\sim1/(\Delta \tau)^{2/3}$. 
Typical numbers for atoms separated by $R\sim 5~\mu\rm m$ and Rydberg principal quantum number $n\sim 75$
are $\Delta\sim 2\pi \times 5 ~\rm MHz$ and $\tau\sim 200 ~\mu\rm s$ leading to errors $E < .01.$

Similar error estimates apply to the above entanglement protocol, but in addition to the imperfect blockade and the atomic spontaneous decay, we must also take into account the undesired interaction between atoms in the Rydberg  $|p\rangle$ transfer states.
The variation in this interaction comes, on the one hand, from considering atoms with all nearest neighbors present relative to atoms at the edge of the ensemble with fewer neighbors and, on the other hand, from the quantum mechanical spreading of the occupancies of the Rydberg state $|p\rangle$ in the time evolving many-atom superposition states.

Returning to Fig. \ref{fig.1}, we expect process ii) to be fast and most sensitive to the blocking interacting $\Delta_{\rm sp}$, but also most sensitive to the $\Delta_{\rm pp}$ shifts, when $\Delta_0=0$. We will treat the spontaneous decay and the imperfect blocking due to finite  $\Delta_{\rm sp}$ as independent errors on each atom. The spontaneous emission error during the transfer between $|0\rangle$ and $|1\rangle$ via $|p\rangle$ is readily determined from the average population of the Rydberg state to be  $E_{\rm se}= \frac{\sqrt2\pi}{4 \Omega_{\rm p} \tau_{\rm p}}.$
The states $|0\rangle$ and $|1\rangle$, coherently coupled to $|p\rangle$ with equal Rabi frequencies $\Omega_p$, can be alternatively treated in the basis of the uncoupled, ``dark" state
 $|d\rangle=(|0\rangle -|1\rangle)/\sqrt{2}$ and the coupled, ``bright"  state $|b\rangle=(|0\rangle +|1\rangle)/\sqrt{2}$ with Rabi frequency $\sqrt{2}\Omega_p$. A $2\pi$ rotation on the $|b\rangle-|p\rangle$  transition yields a sign change on the bright state, equivalent to the desired net $\pi$ rotation between the  $|0\rangle$ and $|1\rangle$ states.
This motivates the definition of $\Omega=\Omega_{\rm p}/\sqrt2$ as  an effective Rabi frequency of the oscillation between $|0\rangle$ and $|1\rangle$, and
in the limit of $\Omega\gg 1/\tau_{\rm p}$ the populations of states $|0\rangle,|1\rangle$ after the transfer pulse (ii) are $P_0=1-P_1$,
\begin{eqnarray} \label{eq.transfer}
P_1&=&\left[\Omega'^2 - \Omega^2  +\Omega^2\cos(\pi\Omega'/\Omega)\right.\nonumber\\
 &-&\Omega'^2\cos(\pi\Delta/2\Omega)\cos(\pi\Omega'/2\Omega)\nonumber\\
&-& \left. \Delta\Omega'\sin(\pi\Delta/2\Omega)\sin(\pi\Omega'/2\Omega)\right] /(2\Omega'^2),
\end{eqnarray}
$\Omega'=\sqrt{4\Omega^2+\Delta^2},$ and $\Delta$ is the interaction induced detuning. The error due to imperfect blocking by the control state $|s\rangle$ is found by inserting  $\Delta\rightarrow \Delta_{\rm sp}\gg \Omega$ and this leads to   $E_{\rm bl}\simeq\frac{\pi^2}{4}\left< \frac{ \Omega^2}{\Delta_{\rm sp}^2}\right>,$ where we assume an average over atom pairs in the ensemble.

Equation (\ref{eq.transfer}) also allows an estimate of the transfer error on each atom scaling as  $E_{\rm tr} \propto \frac{ \Delta_{\rm pp}^2(d)}{\Omega^2}$ due to the $\Delta_{\rm pp}$ interaction terms, when the transfer is not blocked, but we will evaluate this error taking into account the full many-atom correlations in the quantum state. Our task is to determine the effect of the interaction:
\begin{equation}
  V=\sum_{i,j>i} \Delta_{\rm pp}^{ij} (|p_i\rangle \langle p_i|)\otimes (|p_j\rangle \langle p_j|),
\end{equation}
 where $|p_i\rangle \langle p_i|$ is the projection operator of the $i^{th}$ atom on the Rydberg state $|p\rangle$, and $\Delta_{\rm pp}^{ij}$ is the interaction energy for a given $(ij)$ pair of Rydberg excited atoms, depending on their spatial separation. We will determine the effect of this interaction by first order perturbation theory, in the interaction picture with respect to the ideal gate operation due to the Hamiltonian ${\mathcal H}_2$ transferring the atoms between state $|0\rangle$ and $|1\rangle$. That Hamiltonian is readily diagonalized for each atom,
${\mathcal H}_{2i}=\sum_m \omega_m |m_i\rangle \langle m_i|,$
where both the energies $\omega_m$ and the states $|m_i\rangle$ of the $i^{th}$ atom are analytically known (and the same for all atoms), and the corresponding ideal time evolution operator $U_2(t)$ of the entire atomic ensemble is thus also known. The time evolution in the interaction picture, due to the Rydberg interaction is given to first order by the expression:
\begin{eqnarray} \label{eq.pert}
U_I(t)&=
&I-i\sum_{i,j>i} \Delta_{pp}^{ij}
\sum_{m_i,m_j} \sum_{m'_i,m'_j} c_{m_i}c_{m_j}c^*_{m'_i}c^*_{m'_j}\nonumber \\
 &&\times\frac{e^{\imath(\omega_{m_i}+\omega_{m_j}-\omega_{m'_i}-\omega_{m'_j})t}-1}
 {i(\omega_{m_i}+\omega_{m_j}-\omega_{m'_i}-\omega_{m'_j})} \nonumber \\
 &&\times(|m_i\rangle \langle m'_i|)\otimes (|m_j\rangle \langle m'_j|) \otimes \bigotimes_{k \neq (i,j)} I_k,
\end{eqnarray}
where $c_{m_i}=\langle m_i|p_i\rangle$ comes from the expansion of the Rydberg states on the eigenstates of ${\mathcal H}_{2i}$.

The perturbation leads to an erroneous change of the state in the interaction picture, and one readily observes that the squared norm of the erroneous component has the expected $\frac{ \Delta_{\rm pp}^2}{\Omega^2}$ scaling. We have evaluated (\ref{eq.pert}) by summing over all atom pairs and assuming the initial state with all atoms in state $|0\rangle$, and we find for different atom numbers the transfer error $E_{{\rm tr},N}= \alpha_N \cdot \frac{ \Delta_{\rm pp}^2(d)}{\Omega^2}$.
We have carried out calculations with a pair of atoms, separated by the distance $d$, four atoms located at the corners of a square with side length $d$, and 8 atoms located at the corners of a cube with side length $d$. With the $R^{-6}$ scaling of the interaction with distance, we find $(\alpha^{(6)}_2, \alpha^{(6)}_4, \alpha^{(6)}_8)=(0.299,0.72,9.39)$,
while for a distance independent coupling, $(\alpha^{(0)}_2, \alpha^{(0)}_4, \alpha^{(0)}_8)=(0.299,3.82,36.8)$.
The accuracy of the perturbation theory results has  been verified by direct numerical solution of the $N$ atom Schr\"odinger equation for $N\le 6$ .

\begin{figure}[!t]
\centering
\includegraphics[width=8.cm]{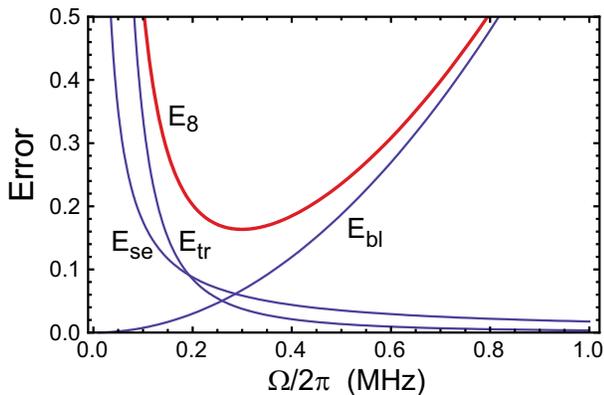}
\vspace{-.5cm}
\caption{(color online) Error of the $N=8$ cat state calculated from Eq. (\ref{eq.error3}) using the parameters of Fig. 2, and $\tau_{\rm p}=57~\mu\rm s$ for $n_{\rm p}=40$.
}
\label{fig.3}
\end{figure}

The approximate cubic growth of the transfer error with the number of atoms qualitatively agrees with an error amplitude on each atom scaling linearly with the number of perturbing atoms. For larger ensembles, distant neighbors do not contribute to the error, and we expect a transition to a linear dependence with $N$.

Adding together our error contributions, we find the total error on the N-atom state,
\begin{equation}
E_N=N\left[ \frac{\pi}{4}\frac{1}{\Omega\tau_p}+\frac{\pi^2}{4}
\left< \frac{\Omega^2}{\Delta_{\rm sp}^2}\right> \right] + \alpha^{(6)}_N
\frac{\Delta_{\rm pp}^2(d)}{\Omega^2}.
\label{eq.error3}
\end{equation}
As can be seen in Fig. \ref{fig.3} the contributions to the error depend in different ways on the Rabi frequency. We find for the case of $N=8$ atoms a minimum error of $E_8=0.16$ at $\Omega/2\pi = 0.30~\rm MHz.$ The 8 atom cat state can thus be prepared with  a fidelity of $\sim 0.84.$

We have also performed the calculation assuming $\Delta_0 \gg \Delta_{\rm pp}(d)$. The $\pi$ pulse in transfer step (ii) leading to the state of Eq. (\ref{eq.psi2}) gives in this case a spontaneous emission error per atom of $E_{\rm se}=\frac{\pi}{2}\frac{1}{\Delta_0\tau_p}$. In the limit of $\Omega\ll \Delta_0$ the population of the states $|0\rangle,|1\rangle$ after the transfer pulse are
$P_0=\cos^2\left({\Omega t_2/2}\right),~P_1=\sin^2\left({\Omega t_2/2}\right)$ with  $\Omega=\Omega_p^2/2(\Delta_0+\Delta_{\rm sp})$ when the $|s\rangle$ state is supposed to block the transition. The blockade error in the target state probability for each atom is  then determined as $E_{\rm bl}= \frac{\pi^2}{4} \left< \frac{\Delta_0^2}{\Delta_{\rm sp}^2}\right> .$ When the transition is not blocked, we shall use our perturbative expression (\ref{eq.pert}), which in the non-resonant case yields an expression of the form $E_{{\rm tr},N}= \beta_N \cdot \frac{ \Delta_{\rm pp}^2(d)}{\Delta_0^2}$. With the same arrangement of 4 and 8 atoms as above we find with the $R^{-6}$  interaction, $(\beta^{(6)}_4,\beta^{(6)}_8)=(15.6, 113)$,
 while for a distance independent coupling, $(\beta^{(0)}_4,\beta^{(0)}_8)=(53.7,308)$.
In this case, the errors add to
\begin{equation}
E_N=N\left[\frac{\pi}{2}\frac{1}{\Delta_0\tau_p}+\frac{\pi^2}{4}\left<
\frac{\Delta_0^2}{\Delta_{\rm sp}^2}
\right> \right]+ \beta_N^{(6)} \frac{\Delta_{\rm pp}^2(d)}{\Delta_0^2}.
\label{eq.error1}
\end{equation}
As the $\beta_N$ coefficients are substantially larger than the corresponding $\alpha_N$ the non-resonant transfer case has a lower fidelity than for the resonant case.

In summary we have presented a technique for preparing multi-atom maximally entangled states using a three step sequence. A detailed analysis of asymmetric Rydberg interactions in Rb atoms shows that 8 atom cat states can be prepared with reasonably high fidelity. Similar results, not presented here, have been found for the case of Cs. Straightforward modifications to these ideas can be used for two-atom CNOT gates that do not require single qubit rotations\cite{Ohlsson2002}, and in
other physical settings where blockade interactions are available
such as Coulomb or Pauli blockade of quantum dots\cite{Friesen2007}, or molecular interactions with superconducting qubits\cite{Tordrup2008}.

This research was supported   by the NSF, ARO, and IARPA.

\end{document}